# Seeing through multimode fibers with real-valued intensity transmission matrices

**TIANRUI ZHAO,1 SEBASTIEN OURSELIN,1 TOM VERCAUTEREN,1 AND WENFENG XIA1,***

*1School of Biomedical Engineering and Imaging Sciences, King's College London, 4th Floor, Lambeth Wing St Thomas' Hospital London, London SE1 7EH, United Kingdom*
*\*wenfeng.xia@kcl.ac.uk*

**Abstract:** Image transmission through multimode optical fibers has been an area of immense interests driven by the demand for miniature endoscopes in biomedicine and higher speed and capacity in telecommunications. Conventionally, a complex-valued transmission matrix is obtained experimentally to link the input and output light fields of a multimode fiber for image retrieval, which complicates the experimental setup and increases the computational complexity. Here, we report a simple and high-speed method for image retrieval based on our demonstration of a pseudo-linearity between the input and output light intensity distributions of multimode fibers. We studied the impact of several key parameters to image retrieval, including image pixel count, fiber core diameter and numerical aperture. We further demonstrated with experiments and numerical simulations that a wide variety of input binary and gray scale images could be faithfully retrieved from the corresponding output speckle patterns. Thus, it promises to be useful for highly miniaturized endoscopy in biomedicine and spatial-mode-division multiplexing in telecommunications.

## 1. Introduction

Multimode fibers (MMFs) have been increasingly attractive for applications in biomedical endoscopy and telecommunications, owing to the capability of transporting images via a large number of transverse optical modes. For biomedical endoscopy, the number of transverse modes in an MMF represents the number of pixels in the images. Compared to multi-core coherent fiber bundles that` are commonly used in biomedical endoscopy, MMFs are significantly more cost-effective, and the effective pixel density in an MMF can be 1-2 orders of magnitudes greater [1-2]. For telecommunications, MMFs are attractive due to the potential of multiplexing data signals within the large number of modes. However, light propagation in MMFs suffers from modal dispersion and mode coupling; for example, when projecting an image pattern onto the proximal end of an MMF, the light field couples into different modes with different propagation constants and thus forms a random-like speckle pattern at the distal end [1-4].

In the last decade, wavefront shaping has been an emerging method for controlling light transport through disordered media such as an MMF [5-32]. A number of research groups [11-12,14-15,23-26,29-31] have demonstrated methods based on the transmission matrix (TM) theory for image transmission through disordered media. In this theory, the optical system of a disordered medium is characterized by a complex-valued matrix, which connects the input and output light fields with transmission constants that have amplitude and phase information. Based on this matrix, both the input and output light fields are divided into orthogonal [11,14,29,30] or spatial frequency modes [1,26,33], whilst a transmission constant linking an input mode to an output mode represents the change of light field during the transport between these two modes. Therefore, the input light field and hence the input intensity pattern can be calculated when the TM and the output light field with phases and amplitudes information are measured. However, conventional cameras are only able to capture light intensity (square of amplitude), so the phase information is usually obtained with holographical methods [1,26,31], which require the use of a complex optical reference arm that can degrade the system stability.

Recently, image retrievals through disordered media using only the intensity information of the output light field were achieved with deep learning [34-41] and model-based methods [42], allowing a simpler optical system compared to those of TM-based methods. With deep learning, a large set of input-output image pairs were used to train a multi-layer neural network for predicting the input images from the output speckles [34-41]. However, the requirements for large training datasets and iterative optimization processes resulted in prolonged time for data acquisition and neural network training. Further, the performance of the trained neural network is likely to depend on similarities between the testing and training datasets. Most recently, a model-based method was used to characterize an MMF [42]. With this method, a set of input images were projected onto one end of an MMF whilst the intensity values of the outputs at the other end were converted into amplitude (square root of the intensity) information with zero phase. With these amplitudes and intensities in the input-output pairs, an iterative algorithm was used to obtain a complex-valued matrix as the inverse TM of the fiber. Different from the deep learning methods, this matrix explicitly linked the input images with the output speckles with a physically-informed model and hence allowed the retrieval of images of complex natural scenes [42]. However, this method also requires large datasets and time-consuming iterative optimization for calculating the complex-valued matrix, which limits its applicability. Thus, a reference-arm-free and high-speed approach for MMF characterization and image retrieval remains highly desired.

In this work, we introduce a method to characterize an MMF with a real-valued intensity transmission matrix (RVITM), which connects the input and output light intensities and hence to retrieve input images from the intensity values of the output speckles. This method is based on an assumption of a pseudo-linearity between the input and output intensity distributions of MMFs under specific conditions as investigated in this study. Importantly, as calculating RVITMs was achieved without a time-consuming iterative process or a large training dataset, this reference-arm-free method allowed a high-speed fiber characterization and image retrieval performed within ~16 s (~8.2 s for data acquisition and ~7.8 s for RVITM calculation) and 0.01 s for 1024-pixel images, respectively, using a desktop PC (Intel i7 8700, 3.2 GHz) and an unoptimized software implementation. Our approach thus could be useful for several applications in biomedical endoscopy and telecommunications.

## 2. Methods and Materials

*2.1 RVITM and image retrieval*

Image retrieval via a RVITM comprised a characterization step and an image reconstruction step as illustrated in Fig. 1 (a) and (b). In the characterization step, as the DMD provides only binary modulations between '1' and '0', a Hadamard matrix $H \in (-1, +1)$ with a size of N by N (notably, N is the power of 2), was converted into two binary matrices $H_1$ and $H_2$ via the formulas $H_1 = (H+1)/2$, and $H_2 = (-H+1)/2$, respectively. Each column of a binary matrix [$H_1$, $H_2$] was then converted to a square matrix as a binary pattern displayed on the DMD, resulting in a total number of 2N binary Hadamard patterns in the input set (the pixel count of each DMD pattern was N). Here we modelled a RVITM that connected the intensities of the input patterns and the output speckles as:

$$\begin{bmatrix} I_1^1 & \cdots & I_1^p \\ \vdots & \ddots & \vdots \\ I_m^1 & \cdots & I_m^p \end{bmatrix} = RVITM \cdot [H_1, H_2],$$
(1)

where $I_{pm}$ represented the intensity value at the $m_{th}$ pixel in the $p_{th}$ output speckle. As all the values in the first binary Hadamard pattern were '1', Eq. 1 was further expressed to arrive at a Hadamard matrix [H, -H] ∈ (-1, +1) on the right-hand side as:

$$\begin{bmatrix} 2I_1^1 - I_1^1 & \cdots & 2I_1^p - I_1^1 \\ \vdots & \ddots & \vdots \\ 2I_m^1 - I_m^1 & \cdots & 2I_m^p - I_m^1 \end{bmatrix} = RVITM \cdot [2H_1 - 1, 2H_2 - 1] = RVITM \cdot [H, -H],$$
(2)

so that the RVITM can be calculated as:

$$RVITM = \begin{bmatrix} 2I_1^1 - I_1^1 & \cdots & 2I_1^p - I_1^1 \\ \vdots & \ddots & \vdots \\ 2I_m^1 - I_m^1 & \cdots & 2I_m^p - I_m^1 \end{bmatrix} \cdot [H, -H]^T,$$
(3)

where [H, -H] T is the transpose of [H, -H] and [H, -H] T = [H, -H] -1 owing to the Hadamard matrix properties.

In the second step, the RVITM was used to reconstruct input images from the output speckles using linear inversion as:

$$I_{image} = RVITM^+ \cdot I_{out} \approx I_{in},$$
(4)

where $I_{image}$ is the intensity distribution of retrieved pattern, $I_{out}$ is the intensity distribution of output speckle measured by the camera. Here, an approximation symbol is used to highlight the approximate linear relationship between the input and output intensities as a pseudo-linearity, and RVITM+ represents the pseudo-inverse of RVITM.

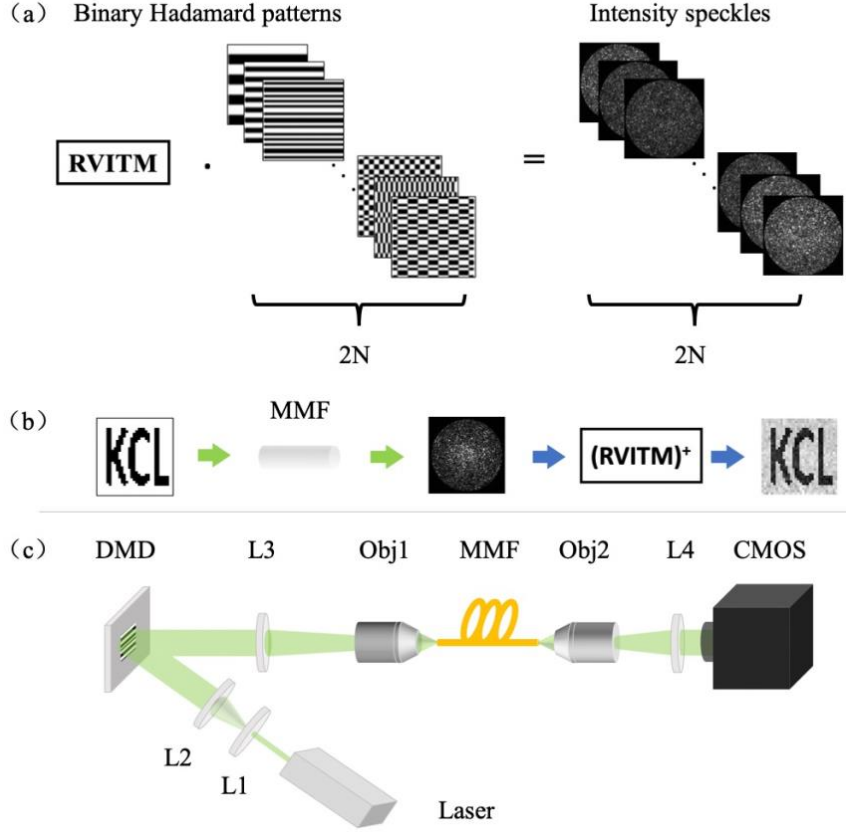

Fig. 1. Schematic illustrations of the principle and experimental setup of image retrieval through a multimode fiber with a real-valued intensity transmission matrix (RVITM). (a) The method of multimode fiber characterization with RVITM. To obtain the RVITM, a set of binary Hadamard patterns are projected onto the proximal end of a multimode fiber with a DMD, whilst intensity values of the speckles at the other end are captured by a camera. (b) The process of retrieving an input image pattern (letters 'KCL') from the corresponding speckle pattern with the calculated RVITM. Notably, RVITM$_+$ represents the pseudo-inverse of RVITM. (c) Schematic diagram of the experimental setup. L1-L4, tube lens; Obj1-Obj2, 20× objectives; DMD, digital micromirrors device; MMF, multimode fiber; CMOS, camera.

## 2.2 Experimental setup

The experimental setup for fiber characterization was illustrated in Fig. 1(c). The intensity of a light beam from a pulsed laser (532 nm, 2 ns, SPOT-10-200-532, Elforlight, Daventry, United Kingdom) was spatially modulated using a DMD (DLP7000, 768×1080 pixels, Texas Instruments, Texas, USA) with a set of binary patterns, and then projected onto the proximal end of a MMF (Ø200 μm, 0.22 NA, 1 m, M122L01, Thorlabs, New Jersey, USA) via a tube lens (AC254-050-A-ML, Thorlabs, New Jersey, USA) and an objective (20×, 0.4 NA, RMS20X, Thorlabs, New Jersey, USA). After magnification with an objective (20×, 0.4 NA, RMS20X, Thorlabs, New Jersey, USA) and a tube lens (AC254-0100-A-ML, Thorlabs, New Jersey, USA), the intensities of the output speckles at the other end were captured by a CMOS camera (C11440-22CU01, Hamamatsu Photonics, Shizuoka, Japan).

*2.3 Experimental investigation*

Several experiments were performed to investigate the conditions of the pseudo-linearity and the impact of several fiber parameters to the fidelity of image retrieval. First, to study the relationship between the intensity distribution of the input image to the image retrieval performance, the number of switched 'ON' micromirrors (J) was varied from 32 to 1024 while the total number of input pixels remained constant (1024). With each J, 64 different input patterns were generated by switching 'ON' the micromirrors at J random positions. A step-index MMF (Table 1, Fiber-200-0.22) was used for this experiment. Second, to study the impact of the total pixel count of input pattern (N), N was varied from 8×8, 16×16, 32×32 to 64×64 in the fiber characterization step. This was realized by combining different numbers of micromirrors on the DMD as one macro-pixel while keeping the number of illuminated micromirrors in the DMD the same (64×64). For example, when 2×2 micromirrors were combined as one input macro-pixel to be switched 'ON' or 'OFF' simultaneously, the input pixel count was 32×32 = 1024. Thus, the resulting RVITM had a size of 1024×M (M is the output pixel count). After characterization, 64 random binary patterns were projected onto the DMD with the same illumination area that covered 64×64 micromirrors as the input ground truths. As such, although the same set of input patterns were used, the pixel counts of the ground truths and the retrieved images varied with N by changing the size of each macro-pixel from 1×1 to 8×8. Third, to study the impact of the numbers of supported transverse modes of MMFs, three different MMFs (as listed in Table 1) were tested. Finally, to demonstrate the impact of the variability of input images, binary images of different types, including handwritten digits [43], schematic plants, animals, Chinese characters and random patterns, were studied for image retrievals through a step-index multimode fiber (Table 1, Fiber-200-0.22). The resulting fidelity of image retrieval was indicated by the correlation coefficients between the retrieved images and the corresponding ground truths. To further improve the fidelity for retrieving binary images, the retrieved images were further binarized using thresholds determined based on their histograms. Here each retrieved binary image exhibited two peaks with intensity values in the vicinities of '0' and '1' in the histogram, and the intensity value corresponded to the valley bottom between the two peaks was chosen as a threshold.

**Table 1. Multimode fibers used for the image retrieval experiments.**

| Fibers | Core diameter (μm) | Cladding diameter (μm) | Numerical aperture | Length (m) | Mode count* | Output pixels count |
|---|---|---|---|---|---|---|
| Fiber-105-0.22 | 105 | 125 | 0.22 | 1 | 9304 | 200×200 |
| Fiber-200-0.22 | 200 | 220 | 0.22 | 1 | 33756 | 360×360 |
| Fiber-200-0.50 | 200 | 220 | 0.50 | 1 | 174360 | 500×500 |

* The mode count is calculated via $N_{mode} = (\pi dNA/\lambda)_2/2$ [33], where d is the diameter of the fiber, λ is the wavelength of the light.

*2.4 Numerical simulations*

As the DMD with our method can provide only binary amplitude modulation, numerical simulations were performed with a custom Matlab program to further study the capability of this RVITM-based method for retrieval of grayscale images. A random, complex-valued TM with 64×64 input pixels and 200×200 output pixels was generated with phases and amplitudes of the TM obeying a uniform and a Rayleigh probability density function between 0 and $2\pi$, and 0 and 1, respectively [5-7]. Output speckle patterns were then calculated using the formula $I_m = |\sum_{n=N} t_{mn} E_n|^2$, where $I_m$ is the intensity at the $m_{th}$ pixel in the output speckle, $t_{mn}$ is the complex-valued transmission constant in the generated TM [5,29], $E_n$ is the light field at the input patterns which was either 1 or 0. Fiber characterization and image reconstruction were performed using the methods presented in Section 2.1; Characterization of the RVITM using the simulated TM was implemented via Eq. 1-3. To evaluate the performance of the image retrieval, two representative binary images and 3 greyscale images (intensity scaled between 0 – 255) with a pixel count of 64×64 were used as input test images. Retrieved images were achieved via Eq. 4.

## 3. Results

The impact of J with input images with a constant pixel count (1024) is shown in Fig. 2. The correlation coefficient increased rapidly to ~90% with J from 32 to 384 and remained largely consistent with J in the range of 384 to 896. With J increasing from 896 to 1024, the correlation coefficient declined rapidly, which can be attributed to the loss of low spatial frequency components due to the diffraction of the DMD micromirrors.

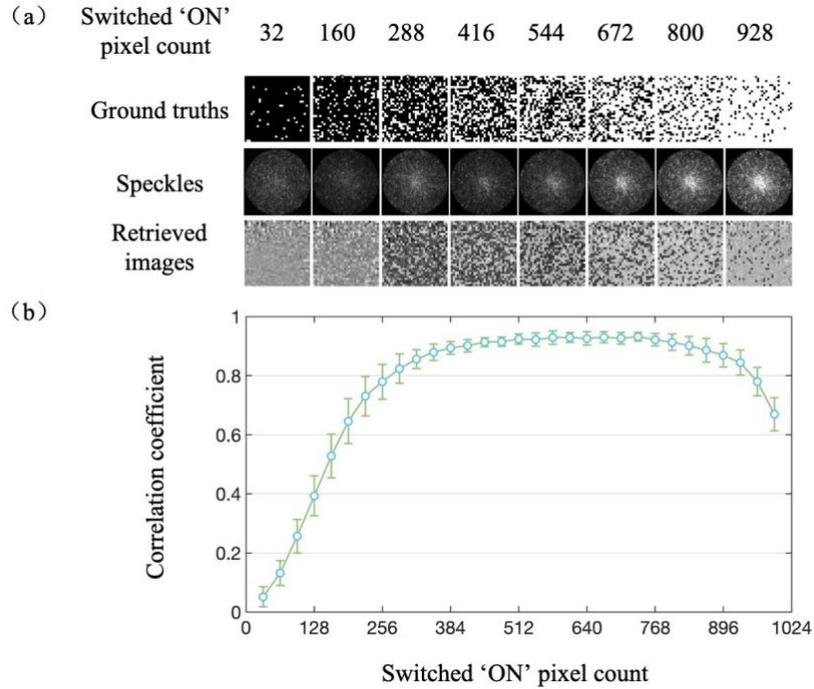

Fig. 2. Impact of switched 'ON' pixels count on the performance of image retrieval. (a) Examples of retrieved 1024-pixel images with varying switched 'ON' pixels count from 32 to 160, 288, 416, 544, 672, 800 and 928 (left to right). (b) The correlation coefficients between the retrieved images and their ground truths versus the switched 'ON' pixels count. Data represent average values across 64 input patterns with the same J, and error bars represent standard deviations.

The impact of the total input pixel count on the performance of image retrieval is shown in Fig. 3. With the same optical fiber, the correlation coefficient declined with the increase of input pixel count. For example, with the MMF having a core diameter of 200 µm, and a NA of 0.22, the correlation coefficients declined from 99.44% to 76.81% with the increase of image pixel count from 8×8 to 64×64. The core diameter and numerical aperture of an MMF also had significant impact on the retrieved image quality, which can be attributed to the impact of the number of supported transverse modes in the MMFs. MMFs with larger numbers of supported transverse modes were able to transmit images with larger input pixel counts. More examples of retrieved images with ground truths with different input pixel counts and through different MMFs can be found in Fig. 7,8 in appendix. Image retrieval of the same image with different level of background noise was shown in Fig. 9 in appendix. The computation time for RVITM estimation increased with the increase of both the input and output pixel counts (N and M). For example, with a desktop PC (Intel i7 8700, 3.2 GHz, 16 GB Memory), when N and M increased from 32×32 and 360×360 (Ø200 µm, NA=0.22, length = 1 m) to 64×64 and 500×500 (Ø200 µm, NA=0.50, length = 1 m), the computation time for RVITM estimation increased from ~8 s to ~240 s, respectively.

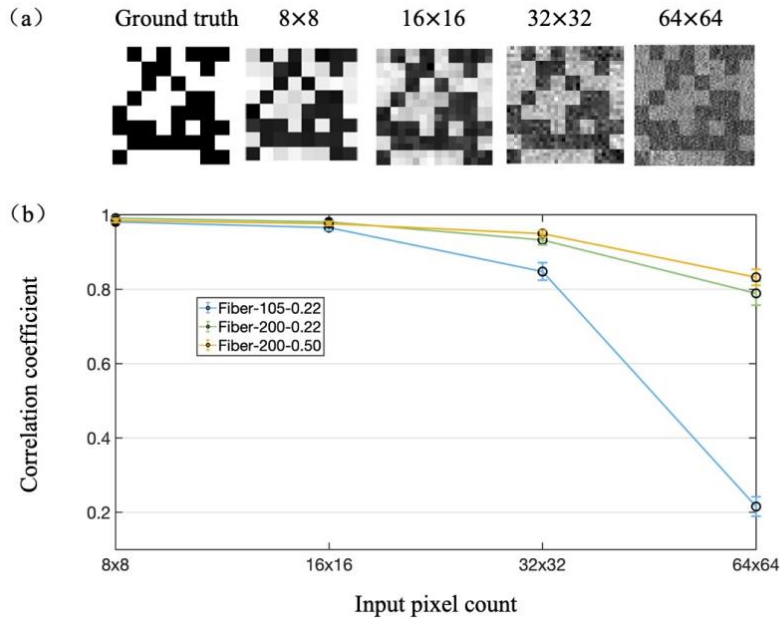

Fig. 3. Impact of input pixel count and number of supported transverse modes in multimode fibers (MMFs). (a) Retrieved images with the same ground truth and different input pixel counts through an MMF with a core diameter of 200 µm and a numerical aperture of 0.22. Retrieved images with the same ground truth and the other two MMFs can be found in Fig. 7 (Appendix). (b) Evolution of image retrieval performance with different MMFs and input pixel counts. Data represent average values across 64 input patterns with the same input pixel count, and error bars represent standard deviations. Fiber-105-0.22, Ø105 µm, NA=0.22, length = 1 m; Fiber-200-0.22, Ø200 µm, NA=0.22, length = 1 m; Fiber-200-0.50, Ø200 µm, NA=0.50, length = 1 m.

As shown in Fig. 4, the fidelity of the retrieved images was weakly dependent on the input patterns: correlation coefficients between the retrieved images and the input ground truths varied from 91.76% for a handwritten digit to 97.62% for a random binary pattern. With a higher number of input pixels (64×64), the correlation coefficient varied from 74.84% for the handwritten digit to 90.91% for the random binary pattern, respectively (Fig. 5). Two videos showing retrieved images from a series of output speckles through the same fiber are shown in Visualizations 1, 2. After binarization, the correlation coefficient increased for all type of input images and with the majority of the input images, the correlation coefficients were greater than 99%.

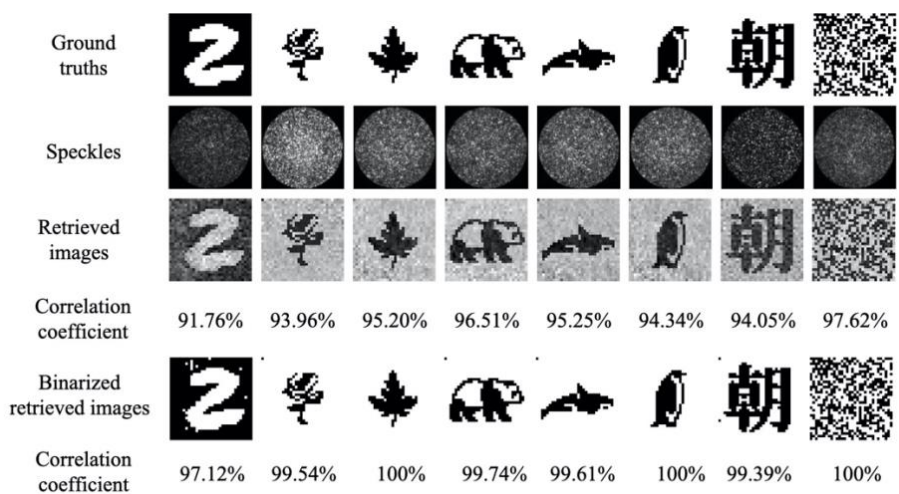

Fig. 4. Image retrieval results with different types of inputs with an input pixel count of 32×32. Retrieved images from corresponding output speckle patterns were obtained from a step-index multimode fiber with a diameter of 200 μm, and numerical aperture of 0.22, and compared to their ground truths. The input pixel counts were 32×32. More examples of image retrievals of handwritten digits can be found in Fig. 8 in appendix.

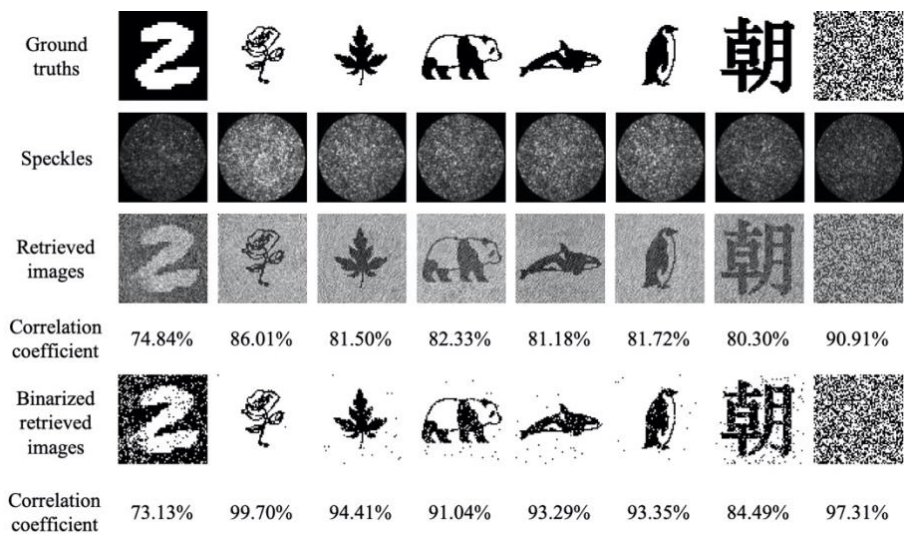

Fig. 5. Image retrieval results with different types of inputs with an input pixel count of 64×64. Retrieved images from corresponding output speckle patterns were obtained from a step-index multimode fiber with a diameter of 200 μm, and numerical aperture of 0.22, and compared to their ground truths.

With numerical simulations, all the binary and grayscale images were retrieved with correlation coefficients greater than 94% (Fig. 6). However, the correlation coefficients for binary images were slightly higher than those for grayscale images.

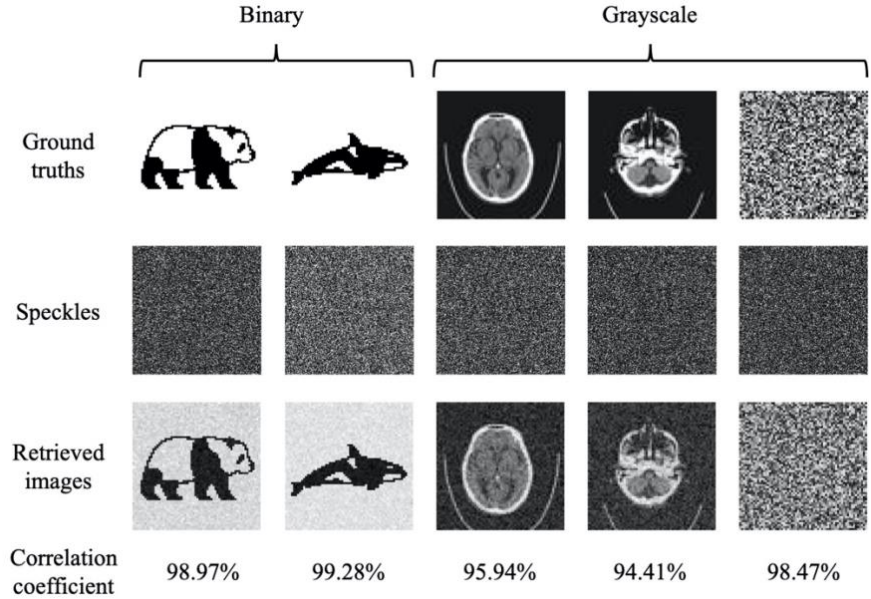

Fig. 6. Numerical simulations of image retrievals with binary and grayscale images. The grayscale head CT images (ground truths) were adapted from the MedNist dataset [44]. Each input ground truth and retrieved image comprised 64×64 pixels while each speckle pattern comprised 200×200 pixels.

## 4. Discussion

In this work, we reported, for the first time to our knowledge, a reference-arm-free and high-speed method to calculate a RVITM that links the input and output intensity distributions of an MMF based on an assumption of a pseudo-linearity between input and output intensities. We studied the conditions of the pseudo-linearity experimentally by investigating the impact of several key parameters to the performance of image retrieval including image pixel count, fiber core diameter and numerical aperture. The results suggest that the fidelity of the image retrieval is a function of the percentage of "1"-valued input pixels to the total number of input pixels. One possible explanation for this dependency is that the characterization input patterns admits a sparse representation of images with 50% pixels 'ON' and so the linear approximation is more suitable for retrieving input images with a similar intensity distribution as those used in characterization. Furthermore, it is evidenced that retrieving input images with a larger number of pixels requires optical fibers supporting more transverse modes. This could be interpreted as when more modes are supported by the fiber, each basis function might tend to use a less overlapping selection of those modes and therefore the optical system has a better linearity. We further investigated the pseudo-linearity and subsequent image retrieval with numerical simulations. The results confirmed the existence of the pseudo-linearity with the numerical models used in the simulations, which suggests that the pseudo-linearity holds true under the conditions of a complex-valued TM with random phases and amplitudes obeying uniform and Gaussian distributions, respectively. It also suggests that our RVITM-based method could be applicable for other types of complex-media including optical diffusers and multicore fiber bundles, for which their TMs have a similar phase and amplitude distributions as those in our simulations. However, further investigations are required to better understand for the pseudo-linearity and its dependency on experimental conditions.

In this study, a pulsed laser was used to demonstrate the capability of our RVITM-based method for image retrieval. It is worth noting that we have previously tested image retrieval with the same method and setup using a continuous wave laser (532 nm, 4.5 mW, Thorlabs, New Jersey, USA). The resulting correlation coefficient between retrieved and input images (~ 93%) were only marginally lower compared to the results obtained with the nanosecond-pulsed laser used in this study. This could be attributed to the short length of the fibers (1 m) which limited the temporal spreading of the output speckles to nanosecond-range, comparable to the pulse duration of the laser. Further reducing pulse duration and/or increasing fiber length is expected to increase the temporal spreading of the output speckles that may have a strong impact on the linearity of the system. This, however, requires further investigations.

Compared with other existing methods, our approach has a number of distinct advantages. First, the fiber characterization process with our method requires both a short time for calculation and data acquisition. In contrast, both the training process of deep learning-based methods and model-based algorithms rely on iterative optimization that is usually very time-consuming (several hours). Furthermore, they require large training datasets and hence a relatively long data acquisition time. This is problematic when repeated fiber characterization process is required due to speckle decorrelations. Second, the performance of image retrieval with our method is weakly dependent on the types of input images. In contrast, deep learning-based approaches are most likely to have a better performance for images that belong to similar classes to those of the training datasets [34-41]. Although a recent study [35] has demonstrated that deep learning networks were able to retrieve images through a MMF that did not belong to the class of images used for training, this success was still limited to simple images. Last, the experimental setup with our method is simpler compared to those of holographical methods which usually require optical reference arms that increase the complexity and can degrade the stability of the system. Several algorithms were reported for TM characterization in reference-arm-free setups [14,24,29,30,45-48], however, these works were mainly focused on producing incident wavefronts based on the achieved TM to focus light through fibers or diffusers. The lack of phase information in output speckles made it challenging to retrieve images directly from the measured TM and speckle intensities. As the RVITM relies on the approximation of a linear relationship, the accuracy of image retrieval with our method is likely to be lower than those of TM-based methods. However, our method could benefit from a higher temporal stability owing to a reference-arm-free setup, as such further studies are required to compare the performance of RVITM, TM, model-based and deep learning methods with the same MMFs and the same test images.

Furthermore, similar to other methods, our method also suffers from speckle decorrelations induced by fiber deformations. One possible mitigation solution is to minimize fiber deformation, e.g. the MMF could be integrated within a rigid medical catheter or needle for biomedical applications. Significant progress has been achieved in recent years to tackle this challenge. In 2015, Plöschner et al. reported that with the knowledge of the fiber shape, TM of the fiber can be corrected to compensate deformations-induced speckle decorrelations [1]. TM characterization using reflected light from metasurface reflector stacks coated on the distal fiber tip was also proposed for solving the bending problem [49]. In 2018, Zhao et al. developed a glass-air disordered fiber, which allowed image reconstruction from output speckles with a deep learning-based method even when the fiber was bent by $90_o$ after characterization [41].

In this paper, the capability of retrieving binary images was demonstrated with experiments and numerical simulations, while the capability of retrieving grayscale images was only demonstrated with numerical simulations. This is due to the DMD used in this work only allowed binary amplitude modulations. In future studies, a spatial light modulator such as the ones in previous studies [34,42] or advanced methods with a DMD [50,51] that provides grayscale amplitude modulations will be used to study the capability of our method for retrieving grayscale images. The potential of a scanner-free endoscopy system based on a single MMF will be exploited, in which the same MMF could be used for both delivering illumination light and collecting reflected light from tissue target similar as that reported by Choi et al. [31]. A pre-characterized RVITM could then be used for recovering tissue images from the reflected speckles captured by a camera at the proximal end of the fiber. The potential of a scanner-free endoscopy system based on a single MMF will be exploited. Based on the results presented in this work, the field-of-view and spatial resolution of the retrieved images depends on the core diameter and NA of the fiber. For example, using a MMF with a core diameter of 200 μm and a NA of 0.22, the highest spatial resolution that could be achieved with the proposed method can be estimated as 200 μm / $64\sqrt{2}$ = ~2.2 μm. MMFs with larger NAs could be used to further improve the spatial resolution.

## 5. Conclusion

In conclusion, we developed a method to measure the light intensity transmission characteristics for an MMF with a RVITM, and with this, light intensity distributions at the proximal end of the fiber can be reconstructed from the measured light intensity distributions at the distal end. This method enables a high-speed characterization of MMFs with a simple measurement setup and thus could be useful for several applications in biomedical endoscopy and telecommunications.

## Appendix

1. Additional image retrieval examples

Additional examples of image retrievals through different multimode fibers, and with different input patterns are shown in Figs 7 and 8, respectively.

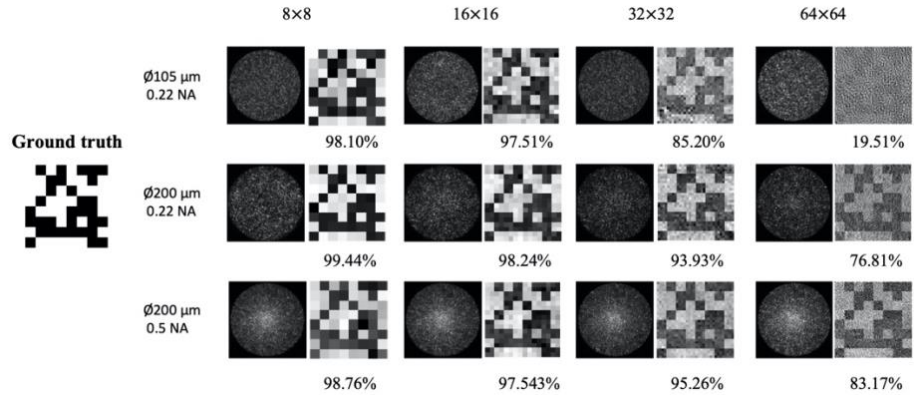

Fig. 7. The results of image retrievals through three different fibers with the same ground truth and varying total pixel counts. Top; Ø105 μm, NA=0.22, length = 1 m; Middle; Ø200 μm, NA=0.22, length = 1 m; Bottom; Ø200 μm, NA=0.50, length = 1 m.

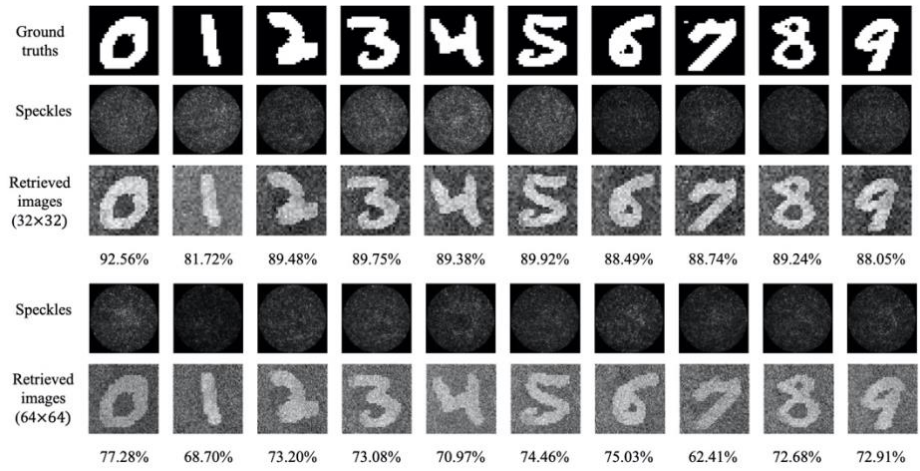

Fig. 8. Some examples of image retrievals through a multimode fiber with handwritten digits with different total pixel counts. The fiber has a core diameter of 200 μm, a NA of 0.22, and a length of 1 m.

2. Noise performance

To study the impact of noise to the performance of the RVITM-based image retrieval method, different levels of simulated background noise were added to experimentally obtained output speckles as $I_{output}' = I_{output} + noise$, and $noise = f \cdot u(I_{output}) \cdot \chi$, where $u(I_{output})$ refers to the mean value of the measured output speckles, and $\chi$ is a matrix with the same size as the output speckles and comprises random values that obey a normal distribution with a mean value of 1 and a standard deviation of 1. f is a parameter that was varied from 0 to 5 resulting in $I_{output}'$ with signal-to-noise ratio (SNR) varied from 1 to 0.2. Here we define the SNR as the ratio of the average value of $I_{output}$ to the standard deviation of the noise. Retrieved images for different SNR levels are shown in Fig. 9. While the correlation coefficient declined with the decrease of SNR, high fidelity images were achieved even when there was a strong background noise.

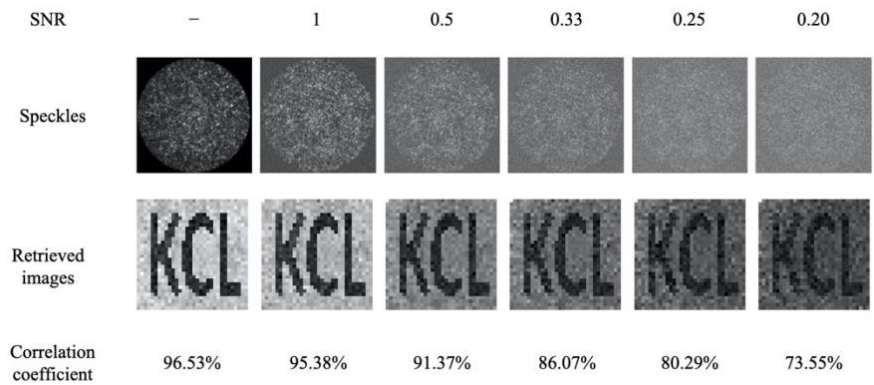

Fig. 9. Impact of noise to the performance of the RVITM-based image retrieval method.

**Funding**

This work was supported by the Wellcome Trust [203148/Z/16/Z; WT101957] and the Engineering and Physical Sciences Research Council (EPSRC) [NS/A000049/1; NS/A000027/1]. T.V. is supported by a Medtronic / Royal Academy of Engineering Research Chair [RCSRF1819\7\34].

**Disclosures**

The authors declare that there are no conflicts of interest. T.V. holds shares from Mauna Kea Technologies, Paris, France, which, however, did not support this work.

**References**

1. M. Plöschner, T. Tyc, and T. Čižmár, "Seeing through chaos in multimode fibers," Nat. Photonics 9(8), 529-535 (2015).
2. E. R. Andresen, S. Sivankutty, V. Tsvirkun, G. Bouwmans, and H. Rigneault, "Ultrathin endoscopes based on multicore fibers and adaptive optics: a status review and perspectives," J. Biomed. Opt. 21, 121506-121506 (2016).
3. W. Xiong, C. Hsu, and H. Cao, "Long-range spatio-temporal correlations in multimode fibers for pulse delivery," Nat. Commun. 10, 1-7 (2019).
4. W. Xiong, P. Ambichl, Y. Bromberg, B. Redding, S. Rotter, and H. Cao, "Principal modes in multimode fibers: exploring the crossover from weak to strong mode coupling," Opt. Express 25(3), 2709-2724 (2017).
5. I. Vellekoop and A. Mosk, "Focusing coherent light through opaque strongly scattering media," Opt. Lett. 32, 2309-2311 (2007).
6. S. Rotter and S. Gigan, "Light fields in complex media: mesoscopic scattering meets wave control," Rev. Mod. Phys. 89, 015005 (2017).
7. J. W. Goodman, Speckle Phenomena in Optics: Theory and Applications (Roberts & Company, 2007).
8. I. Vellekoop, "Feedback-based wavefront shaping," Opt. Express 23, 12189-12206 (2015).
9. A. P. Mosk, A. Lagendijk, G. Lerosey, and M. Fink, "Controlling waves in space and time for imaging and focusing in complex media," Nat. Photonics 6, 283-292 (2012).
10. T. Čižmár, M. Mazilu, and K. Dholakia, "In situ wavefront correction and its application to micromanipulation," Nat. Photonics 4, 388-394 (2010).
11. T. Čižmár and K. Dholakia, "Shaping the light transmission through a multimode optical fiber: complex transformation analysis and applications in biophotonics," Opt. Express 19, 18871-18884 (2011).
12. T. Cižmár and K. Dholakia, "Exploiting multimode waveguides for pure fiber-based imaging," Nat. Commun. 3, 1027 (2012).
13. A. Boniface, B. Blochet, J. Dong and S. Gigan, "Noninvasive light focusing in scattering media using speckle variance optimization," Optica 6(11), 1381-1385 (2019).
14. A. Dremeau, A. Liutkus, D. Martina, O. Katz, C. Schülke, F. Krzakala, S. Gigan, and L. Daudet, "Reference-less measurement of the transmission matrix of a highly scattering material using a dmd and phase retrieval techniques," Opt. Express 23, 11898-11911 (2015).
15. A. Boniface, M. Mounaix, B. Blochet, R. Piestun, and S. Gigan, "Transmission-matrix-based point-spread-function engineering through a complex medium," Optica 4, 54-59 (2017).
16. R. Horstmeyer, H. Ruan and C. Yang, "Guidestar-assisted wavefront-shaping methods for focusing light into biological tissue," Nat. photonics 9(9), 563 (2015).
17. D. Wang, E. H. Zhou, J. Brake, H. Ruan, M. Jang, and C. Yang, "Focusing through dynamic tissue with millisecond digital optical phase conjugation," Optica 2, 728-735 (2015).
18. O. Katz, E. Small, Y. Bromberg, and Y. Silberberg, "Focusing and compression of ultrashort pulses through scattering media," Nat. Photonics 5(6), 372-377 (2011).
19. U. Weiss and O. Katz, "Two-photon lensless micro-endoscopy with in-situ wavefront correction," Opt. Express 26, 28808-2881722 (2018).
20. P. Lai, L. Wang, J. W. Tay, and L. V. Wang, "Photoacoustically guided wavefront shaping for enhanced optical focusing in scattering media," Nat. Photonics 9(2), 126-132 (2015).
21. Y. Liu, C. Ma, Y. Shen, J. Shi, and L. V. Wang, "Focusing light inside dynamic scattering media with millisecond digital optical phase conjugation," Optica 4, 280-288 (2017).
22. R. Di Leonardo and S. Bianchi, "Hologram transmission through multi-mode optical fibers," Opt. Express 19(1), 247-254 (2011).
23. D. Kim, J. Moon, M. Kim, T. D. Yang, J. Kim, E. Chung, and W. Choi, "Toward a miniature endomicroscope: pixelation-free and diffraction-limited imaging through a fiber bundle," Opt. Lett. 39(7), 1921-1924 (2014).
24. D. B. Conkey, A. M. Caravaca-Aguirre, and R. Piestun, "High-speed scattering medium characterization with application to focusing light through turbid media," Opt. Express 20(2), 1733-1740 (2012).

25. O. Tzang, E. Niv, S. Singh, S. Labouesse, G. Myatt, and R. Piestun, "Wavefront shaping in complex media with a 350 kHz modulator via a 1D-to-2D transform," Nat. Photonics 13(11), 788-793 (2019).
26. D. Loterie, S. Farahi, I. Papadopoulos, A. Goy, D. Psaltis, and C. Moser, "Digital confocal microscopy through a multimode fiber," Opt. Express 23(18), 23845-23858 (2015).
27. I. N. Papadopoulos, S. Farahi, C. Moser, and D. Psaltis, "Focusing and scanning light through a multimode optical fiber using digital phase conjugation," Opt. Express 20, 10583 (2012).
28. I. N. Papadopoulos, S. Farahi, C. Moser, and D. Psaltis, "High-resolution, lensless endoscope based on digital scanning through a multimode optical fiber," Biomed. Opt. Express 4(2), 260-270 (2013).
29. S. Popoff, G. Lerosey, R. Carminati, M. Fink, A. Boccara, and S. Gigan, "Measuring the transmission matrix in optics: an approach to the study and control of light propagation in disordered media," Phys. Rev. Lett. 104, 100601 (2010).
30. S. M. Popoff, G. Lerosey, M. Fink, A. C. Boccara, and S. Gigan, "Image transmission through an opaque material," Nat Commun. 1(6), 81 (2010).
31. Y. Choi, C. Yoon, M. Kim, T. D. Yang, C. Fang-Yen, R. R. Dasari, K. J. Lee, and W. Choi, "Scanner-free and wide-field endoscopic imaging by using a single multimode optical fiber," Phys. Rev. Lett. 109, 203901 (2012).
32. O. Katz, P. Heidmann, M. Fink, and S. Gigan, "Non-invasive single-shot imaging through scattering layers and around corners via speckle correlations," Nat. Photonics 8, 784-790 (2014).
33. S. Turtaev, I. T. Leite, T. Altwegg-Boussac, J. M. P. Pakan, N. L. Rochefort, and T. Čižmár, "High-fidelity multimode fibre-based endoscopy for deep brain in vivo imaging," Light Sci. Appl. 7(1), 92 (2018).
34. N. Borhani, E. Kakkava, C. Moser, and D. Psaltis, "Learning to see through multimode fibers," Optica 5, 960-966 (2018).
35. B. Rahmani, D. Loterie, G. Konstantinou, D. Psaltis, and C. Moser, "Multimode optical fiber transmission with a deep learning network," Light Sci. Appl. 7(1), 69 (2018).
36. Y. Li, Y. Xue, and L. Tian, "Deep speckle correlation: a deep learning approach towards callable imaging through scattering media," Optica 5(10), 1181-1190 (2018).
37. A. Turpin, I. Vishniakou, and J. d Seelig, "Light scattering control in transmission and reflection with neural networks," Opt. Express 26(23), 30911-30929 (2018).
38. S. Li, M. Deng, J. Lee, A. Sinha, and G. Barbastathis, "Imaging through glass diffusers using densely connected convolutional networks," Optica 5, 803-813 (2018).
39. P. Fan, T. Zhao, and L. Su, "Deep learning the high variability and randomness inside multimode fibers," Opt. Express 27(15), 20241-20258 (2019).
40. M. Lyu, H. Wang, G. Li, S. Zheng, and G. Situ, "Learning-based lensless imaging through optically thick scattering media," Adv. Photon. 1(3), 036002 (2019).
41. J. Zhao, Y. Sun, Z. Zhu, J. E. Antonio-Lopez, R. A. Correa, S. Pang, and A. Schülzgen, "Deep learning imaging through fully-flexible glass- air disordered fiber," ACS Photon. 5, 3930-3935 (2018).
42. P. Caramazza, O. Moran, R. Murray-Smith, and D. Faccio, "Transmission of natural scene images through a multimode fiber," Nat. Commun. 10(1), 2029 (2019).
43. http://yann.lecun.com/exdb/mnist.
44. https://github.com/apolanco3225/Medical-MNIST-Classification.
45. M. N'Gom, T. B. Norris, E. Michielssen, and R. R. Nadakuditi, "Mode control in a multimode fiber through acquiring its transmission matrix from a reference-less optical system," Opt. Lett. 43(3), 419-422 (2018).
46. T. Zhao, L. Deng, W. Wang, D. S. Elson, and L. Su, "Bayes' theorem-based binary algorithm for fast reference-less calibration of a multimode fiber," Opt. Express 26(16), 20368-20378 (2018).
47. X. Tao, D. Bodington, M. Reinig and J. Kubby, "High-speed scanning interferometric focusing by fast measurement of binary transmission matrix for channel demixing," Opt. Express 23(11), 14168-14187 (2015).
48. G. Huang, D. Wu, J. Luo, Y. Huang and Y. Shen, "Retrieving the optical transmission matrix of a multimode fiber using the extended Kalman filter," Opt. Express, 28(7), 9487-9500 (2020).
49. G. S. D. Gordon, M. Gataric, C. Williams, J. Yoon, T. Wilkinson, and S. E. Bohndiek, "Characterising optical fibre transmission matrices using metasurface reflector stacks for lensless imaging without distal access," Phys Rev X, 9(4), 041050 (2019).
50. K. J. Mitchell, S. Turtaev, M. J. Padgett, T. Čižmár, and D. B. Phillips, "High-speed spatial control of the intensity, phase and polarisation of vector beams using a digital micro-mirror device," Opt. Express 24(25), 29269-29282 (2016).
51. S. A. Goorden, J. Bertolotti, and A. P. Mosk, "Superpixel-based spatial amplitude and phase modulation using a digital micromirror device," Opt. Express 22, 17999-18009 (2014).